\documentclass[a4paper,11pt]{article}
\pdfoutput=1 
\usepackage{jcappub}
\usepackage[T1]{fontenc} 
\usepackage{comment}
\usepackage{float}

\title{Ultra-compact Objects of Non-minimally Coupled Dark Matter}

\author[a,b,c]{Francesco Benetti,}
\author[a,b,c,d]{Andrea Lapi,}
\author[a,b,c]{Samuele Silveravalle,}
\author[a,b,c]{Stefano Liberati}

\affiliation[a]{SISSA, Via Bonomea 265, 34136 Trieste, Italy}
\affiliation[b]{Institute for Fundamental Physics of the Universe (IFPU), Via Beirut 2, 34014 Trieste, Italy}
\affiliation[c]{INFN-Sezione di Trieste, via Valerio 2, 34127 Trieste, Italy}
\affiliation[d]{IRA-INAF, Via Gobetti 101, 40129 Bologna, Italy}

\emailAdd{fbenetti@sissa.it, lapi@sissa.it, ssilvera@sissa.it, liberati@sissa.it}

\abstract{In the framework of a collisionless dark matter fluid which is non-minimally coupled to gravity, we investigate the existence and properties of static, spherically symmetric solutions of the general relativistic field equations. We show that the non-minimal coupling originates an (anisotropic) pressure able to counteract gravity and to allow the formation of regular, horizonless ultra-compact objects of dark matter (NMC-UCOs). We then analyze the orbits of massive and massless particles in the gravitational field of NMC-UCOs, providing some specific example and a general discussion in terms of phase portraits. Finally, we study the gravitational lensing effects around NMC-UCOs, and effectively describe these in terms of a pseudo-shadow.}

\begin{document}
\maketitle
\flushbottom

\section{Introduction}\label{sec|intro}

Several astrophysical and cosmological evidences \cite{Allen2011,Rubin1980,Persic1996,Bennet2003,Aver2015,Planck2018,Zhao2022,Scolnic2018,Garrel2022,Mantz2022,Clowe2006,Paraficz2016} have firmly established that baryons only account for $16\%$ of the total matter of the Universe, the rest being in the form of Dark Matter, a component which interacts mainly via gravity and only very weakly via the other known forces. In the standard cosmological paradigm the latter is supposed to have been already non relativistic at the epoch of decoupling of baryons and radiation, and for this reason it is known as Cold Dark Matter (CDM).

The CDM properties had profound consequences on the formation and growth of perturbations: having non relativistic velocity dispersion they did not suffered free streaming damping, contrariwise to e.g. neutrinos; furthermore, being only weakly interactive with the photon fluid, they did not experienced Silk damping either. As a result, small perturbations in the CDM density could start to grow before recombination, at an epoch when baryons were still tightly coupled to the relativistic component thus smoothing out their smallest perturbations. Being so weakly interactive, on the one hand CDM particles could not develop a strong pressure to offset gravity and on the other hand they could not dissipate any angular momentum preventing a runaway collapse to a pointlike attractor. Thus it is believed that, contrariwise to baryons, they could not form static compact configurations, such as dark matter stars or black holes, but rather loosely bound structures such as virialized halos. There CDM particles are supported in equilibrium by their random motions, which provides a pressure-like component in the Vlasov (collisionless Boltzmann) equation proportional to their dispersion velocity.
After decoupling, baryons could fall in the already formed and virialized CDM halos where, as a consequence of many complex processes, were able to form cosmic structures such as galaxies and galaxy systems. Therefore, each visible galaxy is surrounded by an extended Dark Matter halo whose size and mass exceeds the one of luminous matter by a few orders of magnitude.

Although the true nature of CDM is still unknown, many candidates have been proposed, ranging from ultralight particles with masses as small as $10^{-21}\, \rm eV$, as axions emerging from the solution to the strong CP problem \cite{Peccei:1977hh,Duffy:2009ig}, or more generally axion-like particles \cite{Chadha-Day:2021szb}, to Massive Compact Halo Objects (MACHOs) with masses as large as $10^{-11}\,M_\odot$ such as primordial black holes
\cite{Carr:2016drx,Green:2024bam}. 
In this huge mass range encompassing many orders of magnitude, perhaps the most popular DM candidates have been Weakly Interactive Massive Particles (WIMPs): elementary particles with masses in the $10 \, \rm GeV$ to $100 \, \rm TeV$ range that interact through the weak interaction of the standard model (SM). The interest in these kind of particles is twofold: first, the existence of WIMPs is a natural prediction of many beyond-the-SM particle theories, such as Supersymmetry \cite{Jungman:1995df,Bertone:2004pz}; second, the cosmological DM density observed today can be explained naturally by their thermal relic abundance at freeze-out \cite{Gelmini:2010zh}. This remarkable fact is referred to in the literature as the WIMP miracle. Moreover, the existence of WIMPs can be tested directly in experiments, looking for signals in underground detectors produced by the scattering of DM particles off nuclei mediated by the weak interaction. 
For spin-independent interactions, the leading bounds are from one-to multi-tonne-scale liquid noble gas detectors, including XENON1T \cite{XENON1T}, PandaX-4T \cite{PANDAX}, and LZ \cite{LZ}. For DM masses $m_\chi \sim 20-100 \,\rm GeV$, the upper bound on the Dark Matter-nucleon cross section is of the order $\sigma_{\chi N} < 10^{-47} \, \rm cm^2$, slightly above the irreducible neutrinos background, and much smaller than the cross section $\sigma_{\chi \chi} \sim 10^{-36} \, \rm cm^2$ required by the WIMP miracle. The failed observation of signals from direct detection provide the most stringent constraints to date. These results seem to indicate that, at least for the range of mass considered, there is little room left for a WIMP miracle, and that either CDM particles have a very different mass, or they interact with themselves and with the SM ones with very different strengths.

On the other hand, the current constraints on the weak force for CDM particles open the captivating possibility of considering non SM interactions for particles in the WIMP mass range. The addition of such an interaction could provide a mechanism capable of generating the necessary pressure to sustain a stable gravitational equilibrium, thus substantially altering the picture described above in which compact DM configurations are not possible. Nonetheless, we expect these effects to become relevant at sufficiently high densities to avoid contradiction with the standard CDM model on large cosmological scales. An order of magnitude estimation can be found by requiring the mean free path $\lambda_{\rm mfp} = m_\chi/\rho_\chi\sigma_{\chi \chi}$ of the DM particle to be of the same order of the curvature radius (expressed, e.g., by the inverse of the square root of the Ricci scalar) $R_c = c\,(8\pi G\rho_\chi)^{-1/2}$. This leads to the definition of a threshold density 
\begin{equation}
    \rho_{\rm th} = \left(\frac{8\pi G}{c^2}\right)\left(\frac{\sigma_{\chi \chi}}{m_\chi}\right)^{-2},
\end{equation}
above which the interaction starts to have macroscopical gravitational effects. Assuming a mass $m_\chi\sim 100\,\rm GeV$ and a cross section $\sigma_{\chi\chi}\sim 10^{-36}\,\rm cm^2$ leads to a threshold density of $\rho_{\rm th} \sim 10^{2}\,\rm g/cm^3$, much larger than the universal background, and to length scales $\lambda_{\rm mfp}\sim R_c\sim 10^{12}\,\rm cm$, typical of stars or massive compact objects. Nonetheless, we note that such high densities for Dark Matter are not pertaining to current virialized halos, but could have been attained in the early universe or during the halo virialization phase.

Since almost the only certain thing about Dark Matter is that it couples with gravity, a natural choice for this interaction is a non-minimal coupling with the gravitational field, i.e. $\sigma_{\chi\chi}\to\sigma_{\chi g}$. Non-minimal interactions are commonly introduced in the analysis of scalar fields in gravitational settings due to their generation in the renormalization group flow or their presence in scalar-tensor theories \cite{Birrell:1982ix,Brans:1961sx,Gumjudpai:2015vio}, and they can always be recast as self-interactions \cite{Maeda:1988ab}. In addition, sufficiently small values of $\sigma_{\chi g}$ suggest the intriguing possibility that the non-minimal coupling with gravity could have a universal nature, being present also for SM particles, but that for the latter it would be non relevant because its strength is outclassed by the SM interactions. On the other hand, lacking any other stronger interactions, the non-minimal coupling would be effective for CDM particles. 

In this work, however, we consider a model in which the non-minimal coupling acts on the Dark Matter fluid rather than fields, to avoid assumptions on the precise nature of Dark Matter particles. The idea of such a non-minimal coupling is not completely new, and different models have been considered in the literature. In particular, a series of papers \cite{Bettoni11,Bettoni12,Bettoni14,Bettoni15} has investigated the properties of non-minimally coupled fluids, and their effects on the background evolution of the Universe at late times and on the growth of cosmic structures, while another series of papers \cite{Gandolfi21, Gandolfi22,Gandolfi23} has shown that a Newtonian version of their model can naturally explain the appearance of cores in the innermost part of DM halos within dwarf galaxies. Even though the aforementioned models share a similar structure with the one presented in this work, it must be stressed that the scales involved are completely different: around kpc for the cores of DM halos and several Mpc for the growth of cosmological perturbations, as opposed to the scales $10^{12}\,\rm cm$ we expect. As such, while a Newtonian approximation is appropriate for the models discussed previously, the length and velocity scales considered in the present work require a full relativistic treatment. 

The plan of the paper is as follows: in section \ref{sec|NMC} we introduce the specific model under analysis and its equations of motion; in section \ref{sec|solutions} we solve the equations in the case of regular, static and spherically symmetric solutions, and argue that their main properties allow us to interpret them as Ultra Compact Objects (UCOs); in section \ref{sec|orbits} we present the geodesics of UCOs spacetimes, with a particular focus on the appearance of a pseudo-shadow due to extreme lensing effects; finally, in section \ref{sec|discussion} we summarize the results, and discuss the possible astrophysical implications of these solutions.

\section{Non-minimally coupled Dark Matter}\label{sec|NMC}

In the present work we investigate an effective model in which CDM, characterized as a pressureless dust, is non-minimally coupled with the spacetime curvature. In the relativistic action this is realized by adding a term that involves the contraction of the Einstein tensor (representing the local curvature) with the DM stress energy tensor $T^{\mu\nu}_{\rm DM} = \rho\,c^2\,u^{\mu}\,u^{\nu}$, where $u^{\mu}$ is the fluid velocity.
The total action of the theory then reads
\begin{equation}\label{eq|action}
    S_{\rm NMC} = \int_{\mathcal{M}}{\rm d}^4x\; \sqrt{-g}\,\left[\frac{c^4}{16\pi\,G}R + \mathcal{L}_{\rm DM} + \epsilon L^2 G_{\mu\nu}\,T^{\mu\nu}_{\rm DM}\right]\,,
\end{equation}
where the first term is the Einstein-Hilbert Lagrangian and $\mathcal{L}_{\rm DM}$ denotes the DM Lagrangian. As for the non-minimal coupling term, we considered a perturbative-like construction where $T^{\mu\nu}_{\rm DM} \equiv (2/\sqrt{-g})\,\delta\,(\sqrt{-g}\,\mathcal{L}_{\rm DM})/\,\delta g_{\mu\nu}$ is the DM stress-energy tensor of the minimally coupled theory. In \cite{Minazzoli_2012} the authors have shown that for a barotropic fluid the Lagrangian $\mathcal{L} = -\rho\,c^2$ leads to the desired energy momentum of a perfect fluid, provided the conservation of particle number $\nabla_\mu(\rho\,u^\mu)=0$ applies. As we will see below, this equation fails to be valid once the DM becomes non-minimally coupled with gravity, hence it is not clear how to select a suitable DM Lagrangian from which $T^{\mu\nu}_{\rm DM} = \rho\,c^2\,u^{\mu}\,u^{\nu}$ can be derived. This fact, although being problematic at conceptual level, is irrelevant in the present context since the DM Lagrangian does not appear explicitly in the field equations, which we will consider a valid effective description of a non-minimal coupling. 

To complete the interaction Lagrangian we included the characteristic length $L$ required for the correct dimensionality, and $\epsilon = \pm\,1$ which is a dimensionless `polarity' parameter. The length $L$ can be seen as the macroscopical manifestation of the cross section to mass ratio $\sigma_{\chi g}/m_\chi$ and, as we will see in the next section, it will completely determine the scales of the solutions. The `polarity' parameter $\epsilon$ is instead related with the nature of the non-minimal interaction. It has been shown that in a model with many similarities \cite{Gandolfi21}, the choice $\epsilon=-1$ leads to stable cored configurations for DM halos, while the opposite choice leads to very cuspy inner shapes. In the present context of ultra-compact objects we initially considered both possibilities, but found that the case $\epsilon=+1$ always generates singular solutions, with the emergence of wormhole-like effects; as our focus is on regular configurations, we will set $\epsilon=-1$.

The self-interacting nature of dark matter non-minimally coupled with gravity can be seen rigorously through a disformal transformation into the Einstein frame, which turns the action \eqref{eq|action} into 
\begin{equation}\label{eq|actionef}
    S_{\rm NMC} = \int_{\mathcal{M}}{\rm d}^4x\; \sqrt{-\tilde{g}}\,\left[\frac{c^4}{16\pi\,G}\tilde{R} +\mathcal{F}\left(\rho,\tilde{g}_{\mu\nu}\right) +\mathcal{G}\left(\rho\right)\mathcal{L}_{\rm DM}\right]\,,
\end{equation}
that is a minimally coupled fluid with non-trivial self-interactions. A less rigorous, but more intuitive way to understand this self-interaction is by truly employing a perturbative construction with $R=R^{(0)}+\varepsilon R^{(1)}$, $\mathcal{L}_{DM}=\mathcal{L}_{DM}^{(0)}+\varepsilon \mathcal{L}_{DM}^{(1)}$ and $\epsilon L^2=\varepsilon\epsilon L^2$; the action then takes the form
\begin{align}
    S_{\rm NMC}^{(0)} = & \int_{\mathcal{M}}{\rm d}^4x\; \sqrt{-\tilde{g}}\,\left[\frac{c^4}{16\pi\,G}R^{(0)} +\mathcal{L}_{\rm DM}^{(0)}\right]\,,\\
    S_{\rm NMC}^{(1)} = & \int_{\mathcal{M}}{\rm d}^4x\; \sqrt{-\tilde{g}}\,\left[\frac{c^4}{16\pi\,G}R^{(1)} +\mathcal{L}_{\rm DM}^{(1)}+\epsilon L^2 G_{\mu\nu}^{(0)}T^{(0)\,\mu\nu}\right]= \nonumber \\
    = & \int_{\mathcal{M}}{\rm d}^4x\; \sqrt{-\tilde{g}}\,\left[\frac{c^4}{16\pi\,G}R^{(1)} +\mathcal{L}_{\rm DM}^{(1)}+8\pi\,G\,\epsilon L^2 T_{\mu\nu}^{(0)}T^{(0)\,\mu\nu}\right]\,,
\end{align}
where in the last line we used the $0^{\rm th}$-order Einstein equations. While this construction effectively shows the self-interacting nature of the non-minimal coupling, having a $T_{\mu\nu}T^{\mu\nu}$ term in the action, we will consider the action \eqref{eq|action} in its completeness to address non-perturbative effects of this coupling.

\subsection{Equations of motion and effective stress-energy tensor}

The equations of motion can be easily derived by variation of the action \eqref{eq|action}. In doing so it is convenient to recall the expressions for the variation of the Einstein tensor
\begin{align}
 \delta G_{\mu\nu} = & \nabla^{\alpha}\nabla_{(\mu}\,\delta g_{\nu)\alpha}-\frac{1}{2}\Box\,\delta g_{\mu\nu} -\frac{1}{2}g^{\alpha\beta}\nabla_{(\mu}\nabla_{\nu)}\,\delta g_{\alpha\beta} + \nonumber \\
 & - \frac{1}{2}R\,\delta g_{\mu\nu} - \frac{1}{2}\,g_{\mu\nu}\,(R^{\alpha\beta} + g^{\alpha\beta}\,\Box - \nabla^{\alpha}\,\nabla^{\beta})\,\delta g_{\alpha\beta}\,,
\end{align}
and that of the stress energy tensor for a dust fluid as reported in \cite{haghani2023variation}, where it has been shown to be independent on the choice of the particular form of the fluid Lagrangian,
\begin{equation}
    \delta T^{\mu\nu}_{\rm DM} = \frac{1}{2}T^{\mu\nu}_{\rm DM}\,(u^{\alpha}u^{\beta} - g^{\alpha\beta})\,{\delta g_{\alpha\beta}}.
\end{equation} 
Then a straightforward calculation yields the field equations for the gravitational field sourced by the non-minimally coupled DM fluid

\begin{align}\label{eq|fieldeq}
    \frac{c^4}{8\pi\,G}\,G^{\mu\nu}& = T_{\rm DM}^{\mu\nu} + \epsilon L^2\,\left[(G^{\mu\nu} + g^{\mu\nu}\,\Box - \nabla^{(\mu}\nabla^{\nu)})\,T_{\rm DM} - \Box\,T_{\rm DM}^{\mu\nu} + 2\,\nabla^{\alpha}\,\nabla^{(\mu}\,T^{\nu)}_{\alpha} \right. \nonumber \\
    & \left.- g^{\mu\nu}\nabla_{\alpha}\,\nabla_{\beta}\,T^{\alpha\beta}_{\rm DM} - \frac{R}{2}\,(\,T_{\rm DM}^{\mu\nu}-g^{\mu\nu}\,T_{\rm DM}) - \frac{T_{\rm DM}^{\mu\nu}}{T_{\rm DM}}\,R_{\alpha\beta}\,T^{\alpha\beta}_{\rm DM}\right]\, .
\end{align}

Equation \eqref{eq|fieldeq} reveals a very peculiar feature of the present model: gravity is generated not only by the DM energy density, but also by its variation in space and time. Because of this occurrence, the DM stress energy tensor is no longer conserved; in fact, by taking the covariant divergence of equation \eqref{eq|fieldeq} and using the Bianchi identities, one finds

\begin{align}\label{eq|conservation}  
\nabla_{\mu}T^{\mu\nu}_{\rm DM} &+ \epsilon L^2\,\left\{2\nabla_{\mu}(R^{\nu}_{\lambda}\,T^{\lambda\mu}_{\rm DM}) - \,R^{\mu\nu}\nabla_{\mu}\,T_{\rm DM} -\nabla^{\nu}R_{\mu\lambda}\,T^{\mu\lambda}_{\rm DM}\nonumber \right.\\
& \left.- \nabla_{\mu}\left[\,\frac{R}{2}\,(T^{\mu\nu}_{\rm DM}-g^{\mu\nu}\,T_{\rm DM}) + \frac{T^{\mu\nu}_{\rm DM}}{T_{\rm DM}}\,R_{\alpha\beta}\,T^{\alpha\beta}_{\rm DM}\,\right]\right\} = 0\,.
\end{align}

This fact has profound consequences, which can be best understood by introducing the notion of an effective stress energy tensor. Equation \eqref{eq|action} has the form of the Einstein field equations of General Relativity sourced by an effective energy tensor $G^{\mu\nu} = \kappa\,T_{\rm DM, eff}^{\mu\nu}$. Clearly the latter is not that of a dust fluid; instead, it features both a radial and a tangential pressure, and thus it can be written in the most general form as \cite{Cadoni_2020}
\begin{equation}\label{eq|effSET}
 T_{\rm DM, eff}^{\mu\nu} = (\,\rho_{\rm eff} + p_{\perp\,\rm eff }\,)\,u^{\mu}u^{\nu} + p_{\perp\,\rm eff }\,g^{\mu\nu} + (\,p_{\parallel\,\rm eff} - p_{\perp\,\rm eff}\,)\,k^{\mu}k^{\nu}\,,
\end{equation}
where $k_{\mu}$ is a space-like unit vector orthogonal to the fluid velocity, $k_{\mu}k^{\mu} = +1$, $k_{\mu}u^{\mu}=0$, $\rho_{\rm eff}$ is the DM effective energy density, and $p_{\parallel\,\rm eff }$, $p_{\perp\,\rm eff }$ are the pressures parallel and perpendicular to the vector $k_{\mu}$. These quantities have a dynamical nature, meaning that they emerge as the result of the non-minimal coupling of the original DM dust fluid with the gravitational field; as such, they depend on the particular form of the metric solution of the field equations. Once a metric has been specified, the values of the energy density and pressures can be found by comparison of \eqref{eq|fieldeq} with \eqref{eq|effSET}, and are given by the trace and the projections of the effective stress energy tensor on the vectors $u^{\mu}$ and $k^{\mu}$. 
In view of the above, the meaning of Equation \eqref{eq|conservation} is manifest: it expresses the conservation of the effective stress energy tensor $T^{\mu\nu}_{\rm DM, eff}$, while the NMC with gravity prevents the original tensor $T_{\rm DM}^{\mu\nu}$ to be conserved, as an effective pressure component emerges. In particular, the projection of \eqref{eq|conservation} onto the fluid velocity $u^{\mu}$ is a continuity equation for the DM fluid; the projection on surfaces orthogonal to it yields the relativistic Euler equation, where a dynamically-generated force emerges, implying a deviation of the DM fluid from the geodesic motion. 

All that suggests the remarkable possibility that the DM fluid can settle in a hydrostatic equilibrium configuration where, to use a Newtonian language, the pressure gradient balances the gravitational attraction. This motivates us to look for static solutions of equation \eqref{eq|fieldeq} in a spherically symmetric configuration.

\section{Ultra-compact objects of NMC DM}\label{sec|solutions}

We look for a static, spherically symmetric, asymptotically flat and regular solution of equation \eqref{eq|fieldeq}. In terms of the time $t$ measured by a static observer at infinity, the areal coordinate $r$, and the angular coordinates $\theta$ and $\phi$, the ansatz for the metric can be written in full generality as
\begin{equation}\label{eq|ansatz}
   {\rm d}s^2 = -e^{2\Phi(r)}\,c^2{\rm d}t^2 + \left(\,1-\frac{2\,GM(<r)}{c^2\,r}\,\right)^{-1}{\rm d}r^2+ r^2(\,{\rm d}\theta^2 + \sin^2\theta\,{\rm d}\phi^2\,)~.
\end{equation}
The metric depends on two unknown functions, namely the Newtonian potential $\Phi(r)$, and the cumulative mass distribution $M(<r)$. The latter can be identified with the Misner-Sharp mass, that is the combined total energy of the fluid and gravitational field inside a sphere of radius $r$. Another unknown is represented by the density distribution of the DM fluid in the hydrostatic equilibrium configuration $\rho(r)$. Altogether one has to deal with $14$ equations: $10$ field equations and $4$ conservation equations \eqref{eq|conservation}, but the form of the metric and the general covariance of the theory imply that $11$ of them are either identically zero or equivalent to the remaining $3$, thus leaving just $3$ independent equations for $3$ unknowns. In particular, choosing the $tt$ and $rr$ components of \eqref{eq|fieldeq} and the $r$ component of \eqref{eq|conservation}, we obtain a system of first-order ordinary differential equations in $(M,\rho)$ and a trivial equation for $\Phi$. For the numerical integration will be useful to rescale the variables as
\begin{equation}
    r\to L\,x,\qquad M(<r)\to \frac{c^2L}{8\pi\,G}\hat{M}(<x),\qquad\rho(r)\to\frac{c^2}{8\pi\,G\,L^2}\hat{\rho}(x),
\end{equation}
using which we can write the system of equations in $(\hat{M},\hat{\rho})$ as
\begin{equation}\label{UCOsys}
\left\{
\begin{aligned}
    & \hat{M}'(<x) = 4\pi x^2\frac{\hat{\rho}(x)}{1-\hat{\rho}(x)}\,,\\
    & \hat{\rho}'(x) = -\frac{4\pi x^3\hat{\rho}^2(x)+\hat{M}(<x)\,\left[1-\hat{\rho}(x)\right]^2 }{4 x\,\left[4 \pi  x-\hat{M}(<x)\right]\,\left[1-\hat{\rho}(x)\right]}\,. 
\end{aligned}
\right. 
\end{equation}
The equation for the gravitational potential can be directly integrated in terms of $\hat{M}'$ as:
\begin{equation}\label{eq|phir}
    \Phi(x)  = \Phi_0+\ln\left[\frac{x^2}{4\pi\, x^2+\hat{M}'(<x)}\right] = \tilde \Phi_0+\ln\left[1-\hat{\rho}(x)\right]~,
\end{equation}
where $\Phi_0$ (or $\tilde\Phi_0=\Phi_0-\ln 4\pi$) is an integration constant to be set by the boundary conditions (see below). Once the equations \eqref{UCOsys} are solved, the effective energy density and radial pressure are given by
\begin{equation}
    \hat{\rho}_{\text{eff}}(x)=\frac{\hat{\rho}(x)}{1-\hat{\rho}(x)},\qquad\qquad \hat{p}_{r}(x)\equiv \hat{p}_{\parallel\, \text{eff}}(x)=\frac{4\pi x^3\hat{\rho}^2(x)-\hat{M}(<x)\,\left[1-\hat{\rho}(x)\right]^2}{8\pi x^3\,\left[1-\hat{\rho}(x)\right]^2},
\end{equation}
while the tangential pressure has an involved expression which is not particularly instructive. Note that the condition $\hat{\rho}(x)=1$ defines a critical density $\rho_c=c^2/8\pi\,G\,L^2$ for which the radial pressure and the effective energy density diverge.


We have solved equations \eqref{UCOsys} numerically with initial conditions $\hat{M}(0) = 0$, $\hat{\rho}(0) = \hat{\rho}_0$. As it is evident from the first equation, in order for $\hat{M}'$ to be well behaved near the origin $\hat{\rho}_0\in [0,1]$ must apply. We thus obtain a continuous family of solutions differing by the central value of the density, with $\hat{\rho}_0 = 1$ (that is $\rho(0) = \rho_c$) representing the limiting one. We identify the boundary of the compact object $R_b$ with the radius where the effective radial pressure drops to zero, so that no net radial force acts there. For $r>R_b$ we have matched our solutions with the Schwarzschild metric, since the theory must reduce to General Relativity where the DM stress energy tensor vanishes. In particular, the normalization of the potential \eqref{eq|phir} is chosen such that $\Phi(R_b) = \frac{1}{2}\ln\left[1-\frac{2\,GM(<R_b)}{c^2\,R_b}\right]$, and we have set $M(<r) = M(<R_b)$ and $\rho(r)=0$ for $r > R_b$. However, note that the density drops to zero from a finite value at $R_b$; physically this can be interpreted by stating that, after $R_b$, the DM particles can diffuse freely from the equilibrium configuration, thus reducing drastically the value of the local density. An alternative choice would be to define the boundary as the minimum radius $R_\star$ at which the density attains zero, i.e. $\rho(R_\star) = 0$. However, since $R_\star > R_b$, we have $p_r(R_\star) < 0$, so that the DM fluid would feel an attractive pressure at $R_\star$, making its identification as a boundary ill defined. This ambiguity is not present in General Relativity, where the equilibrium configuration of a spherically symmetric star is found by solving the Tolman-Oppenheimer-Volkoff equations: the form of the latter implies that, for any reasonable equation of state of the matter fluid, the density and pressure approach zero at the same radius, making the identification of the boundary straightforward. In our model, since gravity is generated not only by the DM density but also by its radial variations, the equation of state for the effective stress energy tensor is far from trivial and no simultaneous roots of $\rho$ and $p_r$ are present. 

It is interesting to notice that the choice of boundary $R_b$ ensures that the potential $\Phi$ is of class $C^1$, while the derivative of the mass $M'(<r)$ has a discontinuity at the boundary due to the sudden drop of density. We have checked that having instead identified the boundary with $R_\star$ would have resulted in a mass function of class $C^1$ and a discontinuous derivative of the potential $\Phi'$. In particular, it is not possible to make both components of the metric to be of class $C^1$ over the entire domain. However, a discontinuity in the derivative of the potential would have much deeper consequences on the physical structure of the spacetime. The extrinsic curvature $\kappa_{\mu\nu}(r)$ for a surface of constant radius has a discontinuity at the surface $R_\star$, while it remains continuous using the boundary $R_b$; this discontinuity is inherited by the expansion and shear tensors for a congruence of geodesics, which at the physical level means that nearby free-falling particles would perceive a jump in their relative velocities, that is an infinite relative acceleration.
While the choice of boundary $R_b$ removes such divergences, the relative accelerations of nearby geodesics, as well as invariant contractions of curvature tensors, will have a jump at the boundary. These jumps are unavoidable because they result from the drastic change in gravitational interaction caused by a hard boundary on the extent of the non-minimal coupling. In a realistic setting we expect the non-minimal interaction to diminish smoothly in such a way that the energy density and the curvature tensors get regularized. Nonetheless, we accept and interpret these discontinuities as an effective description for a very sharp falloff in the interaction.

\begin{figure}[t!]
    \centering
    \includegraphics[width=\textwidth]{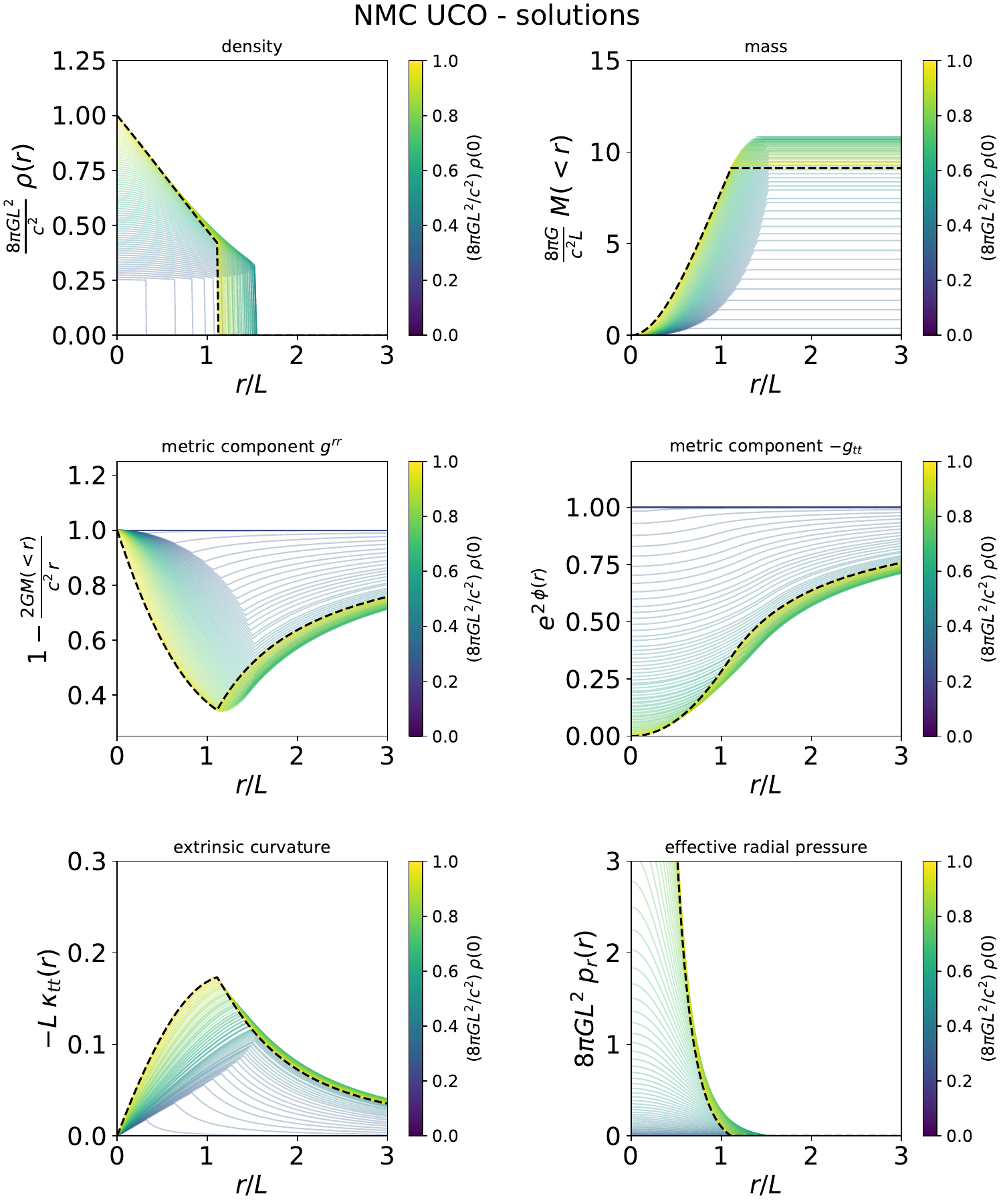}
    \caption{Radial profiles of UCOs solutions: density $\rho(r)$ (top left), cumulative mass $M(<r)$ (top right), metric component $g^{rr}$ (middle left), metric component 
    $-g_{tt}$, extrinsic curvature component $-\kappa_{tt}$ (bottom left), and effective radial pressure $p_r(r)$ (bottom right). The various solutions are parameterized by the allowed values of the central density $\rho(0)$, as specified by the color scale. The dashed black line highlights the limiting solutions featuring the maximal value of the central density.}
    \label{fig|sol}
\end{figure}

\bigskip

\subsection{Properties of the NMC-UCOs}

In Figure \ref{fig|sol} we illustrate the density $\rho(r)$, the cumulative mass $M(<r)$, the metric components $g_{rr}^{-1}$ and $-g_{tt}$, the effective radial pressure $p_{r}(r)$, and the temporal component of the extrinsic curvature $-\kappa_{tt}(r)$ for a surface of constant radius. The colormap refers to the different solutions parametrized by the value of the central density, with the limiting solution with $\hat{\rho}_0 = 1$ highlighted by the dashed black line.

First of all we notice from the top-left panel that there is no solution for $\hat{\rho}_0<1/4$; this minimum central density is the one for which the central radial pressure
\begin{equation}\label{eq|centralpressure}
    \hat{p}_{r,0}=\frac{\hat{\rho}_0\left(4\hat{\rho}_0-1\right)}{6\left(1-\hat{\rho}_0\right)^2}
\end{equation}
is zero and the solution becomes trivial. For smaller central densities it is not possible to define the boundary $R_b$ without introducing a negative energy density for the fluid, which is unphysical for a standard DM framework. Being left only with the possibility of the ill-defined boundary $R_\star$, we argue that central densities $\hat{\rho}_0<1/4$ are not sufficient to sustain a stable solution. The density range of the solutions is therefore extremely small, with central values $\hat{\rho}_0\in\left[0.25,1\right]$ and values at the boundary $\hat{\rho}\left(R_b\right)\in\left[0.25,\sim0.422\right]$. Indeed, these solutions are present only at a precise scale and have a particularly steady density in their interior, suggesting a physical process well-localized in the density domain.

The density radial profile shows that the solutions grow in size with the central density up to $\hat{\rho}_0\sim 0.47$, after which they start shrinking until they reach the limiting solution. On the other hand, the top-right panel shows that the total mass $M\left(<R_b\right)$ reaches a maximum at $\hat{\rho}_0\sim 0.66$, after which it start to decrease; interestingly, the mass of the limiting solution is almost the same as that of the one with the maximum radius. The mass-radius relation for these object will then follow a curve similar to the one of strange stars \cite{Alcock:1986hz}, where at low densities they have small masses and radii increasing concurrently with the central density, while denser objects have larger masses and radii which instead decrease with the central density.

The metric components shown in the middle panels have familiar profiles, with the notable exception for the presence of corners in the radial component $g^{rr}$ due to our choice of the boundary. As discussed before, this behavior at the physical level translates in sudden changes of particle accelerations due to the discontinuous change in density, while the differentiability of the temporal component $g_{tt}$ guarantees that they maintain continuous relative velocities. This is explicitly shown by the continuity of the extrinsic curvature displayed in the bottom-left panel, which indeed indicate that there is no localized stress-energy tensor on the surface. We stress that the UCO solution does not feature any horizon since the $g^{rr}$ component never vanishes.

As the central density approaches the critical value, the temporal component of the metric reduces its central value up to zero for the limiting solution. The vanishing of $g_{tt}$ can be interpreted in a Newtonian sense as an infinite gravitational potential $\Phi$, and in a relativistic sense as an infinite redshift for a photon emitted in the origin and measured at spatial infinity $z=\mathrm{e}^{-\Phi(0)}-1$. In both cases we can state that for the limiting solution the gravitational field in the origin is so intense that an infinite amount of energy is required to leave it. Such a strong gravitational attraction has to be balanced by an equally strong radial pressure, which is indeed shown to diverge for the limiting solution in the bottom-right panel, in agreement with \eqref{eq|centralpressure}. The divergence of the effective pressure, which by definition corresponds to the $rr$ component of the Einstein tensor, indicate that the solution becomes singular. However, this singularity does not have to be interpreted as an exotic naked singularity but rather as a limit for stable solution as the Buchdahl limit in General Relativity.

\begin{figure}[t!]
    \centering
    \includegraphics[scale=0.575]{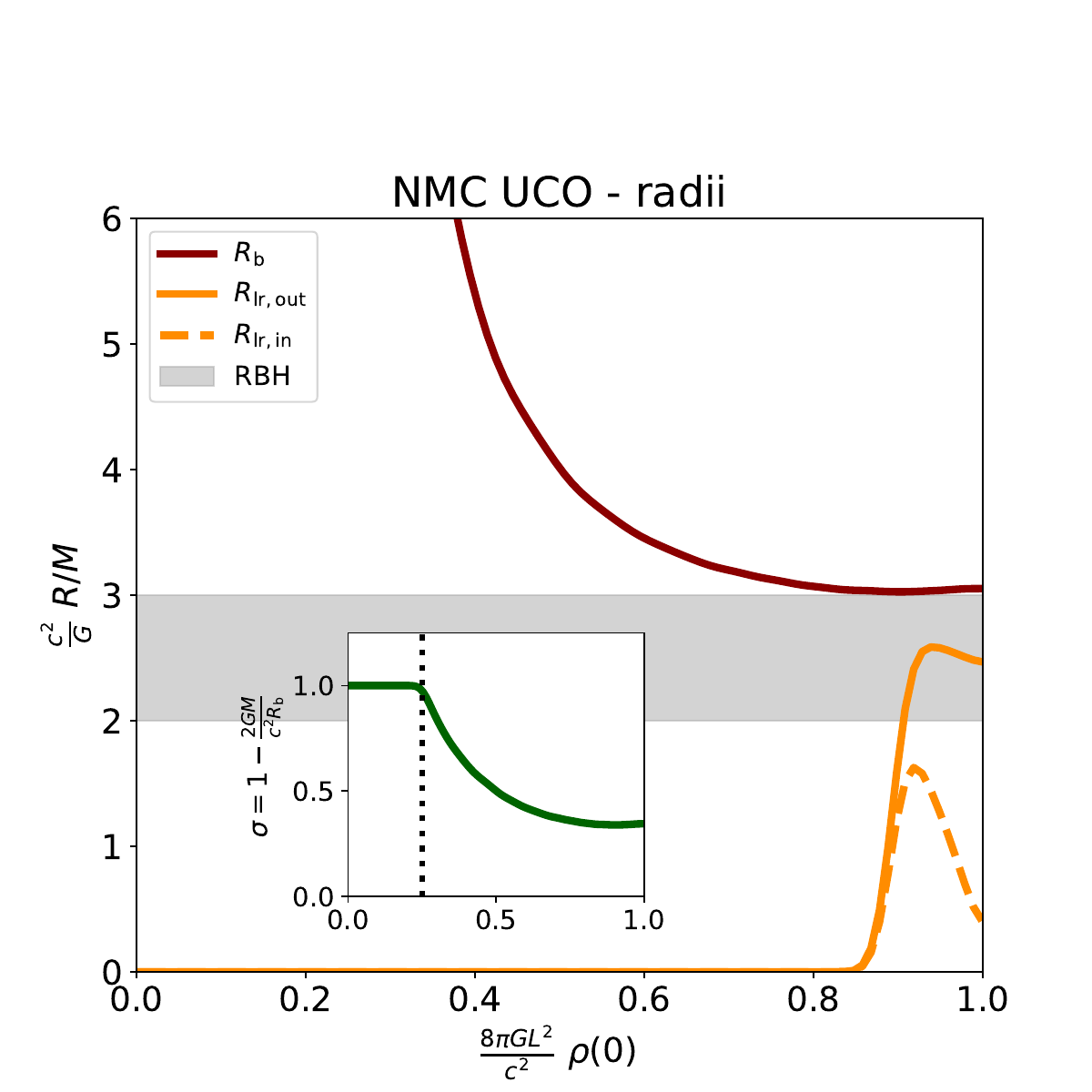}
    \caption{Relevant physical length-scales of UCOs solutions: boundary radius $R_{\rm b}$ (red) and light rings (orange; dashed for the inner and solid for the outer); these are measured in units of total mass $M$ of the UCO, and are plotted as a function of the UCO central density $\rho(0)$ defining the various solutions. The grey shaded area displays the typical size of spherically symmetric regular black hole solutions (e.g., black hole mimickers) in general relativity. The inset displays the compactness parameter $\sigma=1-2 G\, M/c^2\, R_{\rm b}$ (green); the dotted line is the limiting value of the density $\frac{8\pi G\,L^2}{c^2}\, \rho(0)= 0.25$ below which no physical solution exists.}
    \label{fig|radii}
\end{figure}

In any case, the presence of this extreme curvatures indicate that we are in a very strong gravity regime. In Figure \ref{fig|radii} we show some relevant length-scales which indicate that the compactness of the solutions is close to the typical size of a black hole mimicker, especially when one approaches the limiting solution. In particular, in red we show the radius in mass units, which approaches the size of the light ring (i.e. the radius where a circular orbit for photons is allowed) of the Schwarzschild solution $R=3\,GM/c^2$, and in the inset we show the compactness parameter $\sigma=1-2GM\left(<R_b\right)/c^2R_b$ which goes from $1$ at $\hat{\rho}_0=1/4$ to $\sim 1/3$ at $\hat{\rho}_0=1$. Such an high compactness leads to the formation of two light rings, as usual for  very compact regular solutions, of which the larger describes unstable circular photon orbits and the smaller stable ones (we will explain the definition of light ring in the next section). 
Finally, we can conclude that our choice of naming these solutions Ultra-Compact Objects is thus justified.

\section{Orbits around NMC-UCOs}\label{sec|orbits}

To have insight into the physical nature of NMC-UCOs, we analyze both the timelike and null geodesics of their spacetime. As a matter of fact, geodesics analysis is particularly instructive in this case, as we expect both massive and massless test particles to interact with the DM fluid mainly through the gravitational field rather than through the direct non-minimal coupling, therefore following its geodesics. As such their motion is described by the Lagrangian for a free falling particle
\begin{equation}\label{eq|Lagrangian}
    \mathcal{L}_p = \frac{1}{2}g_{\mu\nu}\dot{x}^{\mu}\,\dot{x}^{\nu} = -\frac{1}{2}\,e^{2\Phi(r)}\,c^2\dot{t}^2 + \frac{1}{2}\,\left[\,1-\frac{2\,GM(<r)}{c^2\,r}\,\right]^{-1}\,\dot{r}^2 + \frac{1}{2}\,r^2\,\dot{\phi}^2 \,,
\end{equation}
where a dot denotes derivation with respect to an affine parameter $\lambda$ along the trajectory, and without loss of generality we have assumed the motion to take place on the plane $\theta = \pi/2$. Differentiation of \eqref{eq|Lagrangian} with respect to $\lambda$ yields the geodesic equation \begin{equation}\label{eq|geodesics}
    \ddot{x}^{\mu} +  \Gamma{^\mu_{\alpha\beta}}\,\dot{x}^{\alpha}\dot{x}^{\beta}=0\,,
\end{equation}
implying that the Lagrangian itself is constant along the trajectory. Thus we can restrict to consider the simpler first order equation 
\begin{equation}\label{eq|trajectory}
    \eta = -e^{-2\Phi(r)}\,E^2 + \left(\,1-\frac{2\,GM(<r)}{c^2\,r}\,\right)^{-1}\dot{r}^2 + \frac{j^2}{r^2}\,,
\end{equation}
where the energy $E$, the angular momentum $j$, and the parameter $\eta$ are the first integrals of motion associated to the invariance of the Lagrangian under time translations, rotations in the considered plane, and reparametrizations of the affine parameter along the particle trajectory, i.e.:
\begin{equation}\label{eq|eqconserv}
\begin{aligned}
     E &\equiv -\frac{\partial\mathcal{L}_p}{\partial\dot{t}} = e^{2\Phi}\,\dot{t}\,, \\ 
     j &\equiv \frac{\partial\mathcal{L}_p}{\partial\dot{\phi}} = r^2\,\dot{\phi}  \,, \\ 
    \eta & \equiv g_{\mu\nu}\,\dot{x}^{\mu}\dot{x}^{\nu} \,. 
\end{aligned}
\end{equation}
Specifically, $\eta = 0$ applies to massless particles while, if $\lambda$ is chosen to be the proper time along the trajectory, $\eta = -1$ applies to massive ones.

Equation \eqref{eq|trajectory} can be recast as
\begin{equation}
    \dot{r}^2 = \left(1-\frac{2\,GM(<r)}{c^2\,r}\right)\left(e^{-2\Phi}\,E^2 - \frac{j^2}{r^2}+\eta\right)\,;
\end{equation}
being horizon-less and regular at the origin, our solution has $1-\frac{2\,GM(<r)}{c^2\,r}>0$ in all of space, implying that a photon and a massive particle can propagate only in those regions where 
\begin{equation}\label{eq|allowed}
e^{-2\Phi}E^2 - \frac{j^2}{r^2} \geq -\eta\,,
\end{equation}
with equality corresponding to a turning point with $\dot{r}=0$. In particular, a radial geodesic starting at $r_0$, having $j=0$, will fall through the center of the UCO re-emerging on the opposite side, since $e^{2\Phi(r)}$ is a monotone increasing function of the radius and $E = e^{2\Phi(r_0)}$ holds. On the other hand, if $j\neq 0$ a centrifugal barrier is present, preventing the particle to reach the center. 

To understand what happens at the turning point, we look at the radial component of the geodesic equation \eqref{eq|geodesics} with $\dot{r}=0$:
\begin{equation}\label{turning}
    \ddot{r} = \left(1-\frac{2\,GM(<r)}{c^2\,r}\right)\,\frac{e^{-2\Phi}}{r}\left[(1-r\,\Phi')E^2-\eta e^{2\Phi}\right ]\,.
\end{equation}
This implies that a particle will experience an attractive or a repulsive force at the turning point depending whether $(1-r\,\Phi')E^2-\eta\, e^{2\Phi}$ is below or above $0$ there. This condition is of particular interest in the case of massless particles, where it reduces to whether $r\,\Phi'$ is above or below $1$, with no dependence on the energy. In particular, if there exist one or more radii $R_{\rm lr}$ such that $R_{\rm lr}\,\Phi'(R_{\rm lr}) = 1$, then a photon with energy and angular momentum satisfying $E/|j| = e^{\Phi(R_{\rm lr})}/R_{\rm lr}$ will move on a circular orbit there: a radius with this property is known as a "light ring". 
A detailed  numerical analysis of the solution reveals that for a central value of the normalized density less that $\rho_0 \lesssim 0.85$ no light rings exist, while for higher densities two of them appear: an external one which describes unstable circular orbits and an internal one which describes stable orbits. The stability of the orbits is manifest if we recall that $r\,\Phi'\to 0$ at large distances, and then circular orbits close to the external light ring will experience a repulsive force on the outside and an attractive one on the inside, while close to the internal light ring they will experience repulsive force on the inside and attractive on the outside. For this reason, close to the internal light ring bounded photon orbits are possible. Although the light ring position $R_{\rm lr}<R_{\rm b}$ is internal to the boundary of the UCO, this name is appropriate since the object is made of DM and hence is "transparent" to radiation. 

\begin{figure}[t!]
    \centering
    \includegraphics[width=\textwidth]{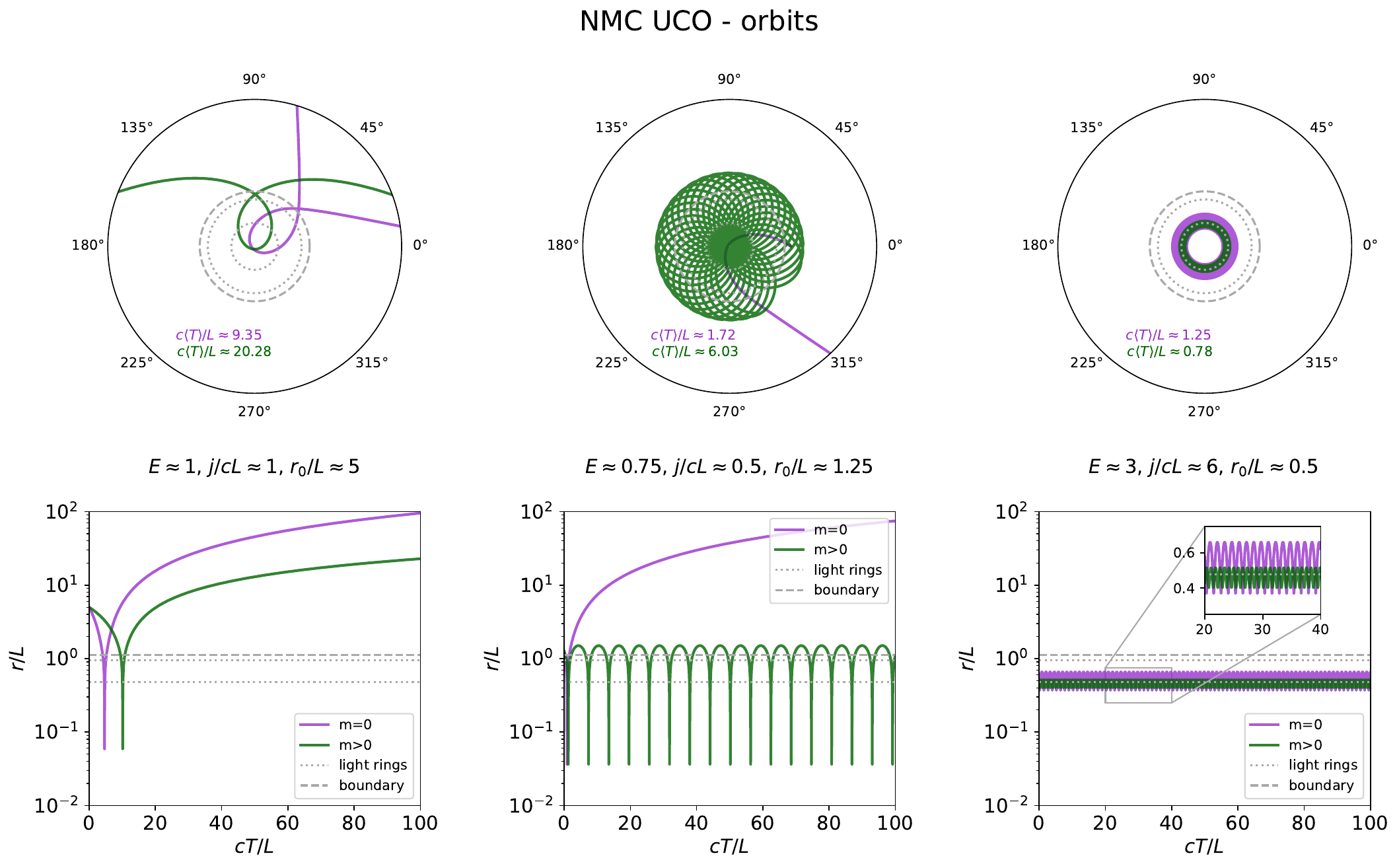}
    \caption{Relevant examples of orbital trajectories for massless (magenta) and massive (green) particles around an UCO solution with central density $\hat{\rho}_0=0.95$. Top diagram show the trajectories in polar coordinates, bottom diagrams illustrate the radial position of the test particle as a function of proper time. The dashed grey lines displays the position of the boundary, dotted grey lines that of the inner and outer light rings. Left diagrams refer to an unbound orbit, middle diagrams to a loose bound orbit for massive particles, and right diagrams to a tigthly bound orbit for both massless and massive particles; the values of the energy parameter $E$, of the angular momentum $j$, of the starting radial coordinate $r_0$ and the average proper period $\langle T\rangle$ of the orbit are reported (polar initial angle is assumed to be null).}
    \label{fig|orbits}
\end{figure}

In Figure \ref{fig|orbits} we showcase some relevant examples of orbital trajectories for massless and massive test particles in the gravitational field of the UCO, obtained by solving the orbital equation
\begin{equation}\label{eq|orbits}
    \left(\frac{dr}{d\phi}\right)^2 = r^2\, \left[1-\frac{2\,GM(<r)}{c^2\,r}\right]\,\left(r^2\,e^{-2\Phi}\,\frac{E^2}{j^2} +\eta\,\frac{r^2}{j^2}-1\right)\,
\end{equation}
with different initial conditions $r_0$, and different values of the energy $E$ and angular momentum $j$. Top diagrams show the trajectories in polar coordinates, magenta for massless and green for massive particles. The average orbital proper time is also reported. Bottom diagrams show instead the radial position of the test particle as a function of proper time (for massive particles), or of the affine parameter along the trajectory (for massless ones). Left diagrams refer to an unbound orbit, middle diagrams to a loose bound orbit for massive particles (note the pericenter precession), and right diagrams to a tightly bound orbit for both kind of particles. To ease the visualization, the boundary of the UCO and the light rings are also displayed in all diagrams as continuous and dashed grey lines, respectively. Note that proper times are of the order of $L/c$ even for massive particles, typical of ultra-relativistic motion; this copes with the very high compactness of the UCO.

\subsection{Orbital portraits}

\begin{figure}[t!]
    \centering
    \includegraphics[width=\textwidth]{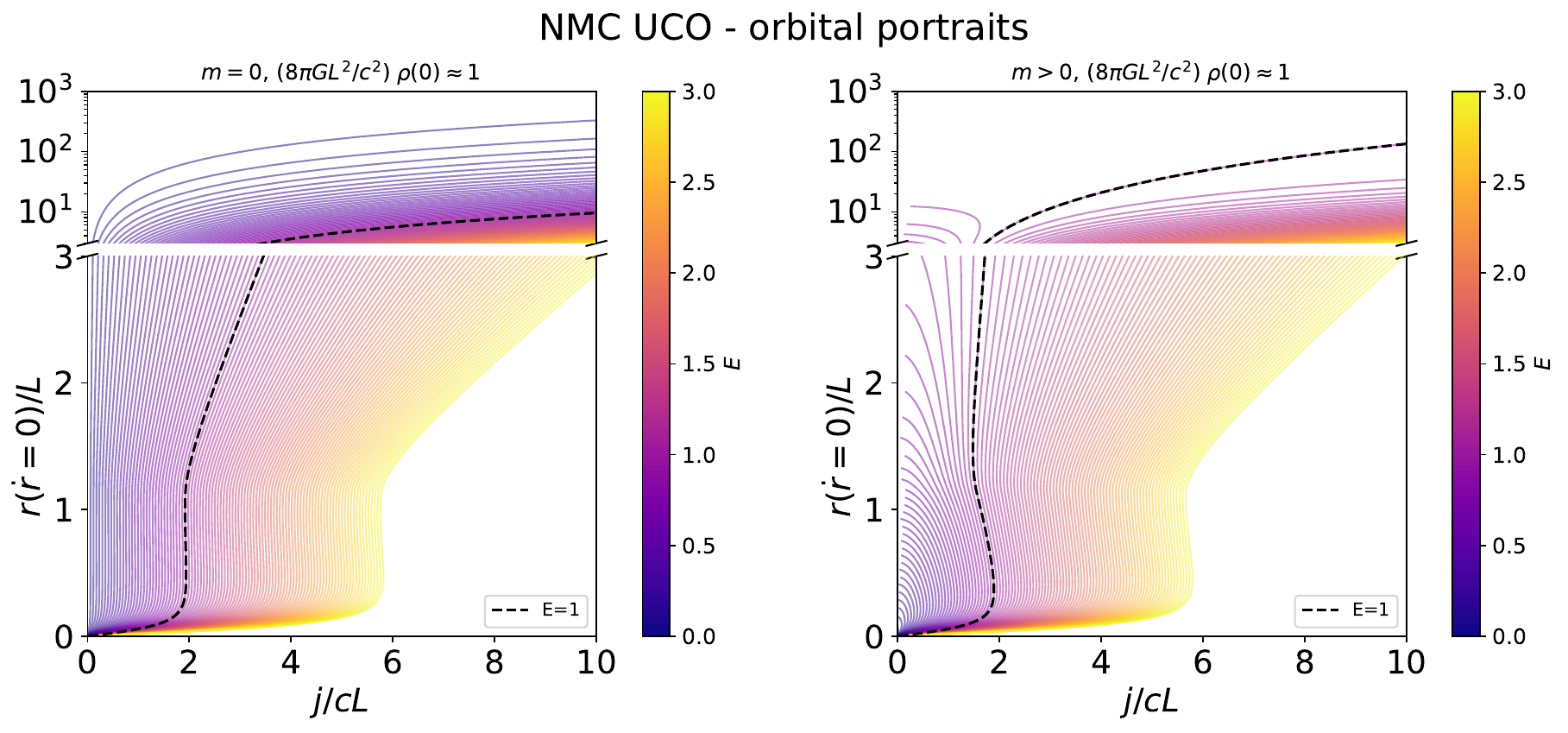}
    \caption{Orbital portraits diagrams for the UCO solution with central density $\hat{\rho_0}=0.95$, slightly below the maximal value. Left panel refers to a massless and right panel to a massive test particle. The radius $r$ at the radial inversion point(s) with $\dot r=0$ is shown as a function of the angular momentum $j$ ($x$-axis) and of the energy parameter $E$ (color-scale) of the test particle. The dashed black line highlights the portraits for $E=1$. Note that in both the panels, the scale of the $y$-axis starts linear at the bottom and then becomes logarithmic at the top.
    }
    \label{fig|orbport}
\end{figure}

While Figure \ref{fig|orbits} makes clear and intuitive what are the possible orbits in UCO spacetimes, it is far from giving a complete picture. Toward this purpose, in Figure \ref{fig|orbport} we show the orbital portraits diagrams for an UCO solution with a central density near the maximum value, high enough to develop light rings. We show the radius $r$  at the radial inversion point(s) with $\dot r=0$, as functions of the angular momentum ($x$-axis) and of the energy parameter (color-scale) of the test particle. Left panel refer to a massless and right panel to a massive test particle. The dashed black line refers to the case $E=1$, appropriate e.g. for a particle starting at rest from infinity. In both panels, the scale of the $y$-axis starts linear at the bottom and then becomes logarithmic at the top.

The number of intersections between a colored curve (corresponding to a given energy $E$) and a vertical line (corresponding to a specific value of the angular momentum $j$) hints to the nature of the possible orbits. In fact, such intersections highlight the different radii $r$ at which motion inversion with $\dot r=0$ can occur. When only one intersection is present (e.g., for both massive and massless particles at high $j$), the particles will reach a minimum radius with zero radial velocity, and then they will be pushed away by the centrifugal barrier and reach spatial infinity. When there are two intersections (e.g., for massive particles with $E\lesssim 1$ and $j/cL\lesssim 2$), a bound orbits is possible between a maximum and minimum radius. Finally, when there are three intersections (this can be clearly seen for massive particles with $E\approx 1$ and $j/cL\approx 1.8$), two kind of motion can occur: if the orbit starts from a large radius, then the particle will encounter the external inversion point and will be pushed away to spatial infinity; contrariwise, if the orbit starts between the two internal turning points, it can be bounded.

\subsection{UCO's pseudo-shadow}

\begin{figure}[t!]
    \centering
    \includegraphics[width=\textwidth]{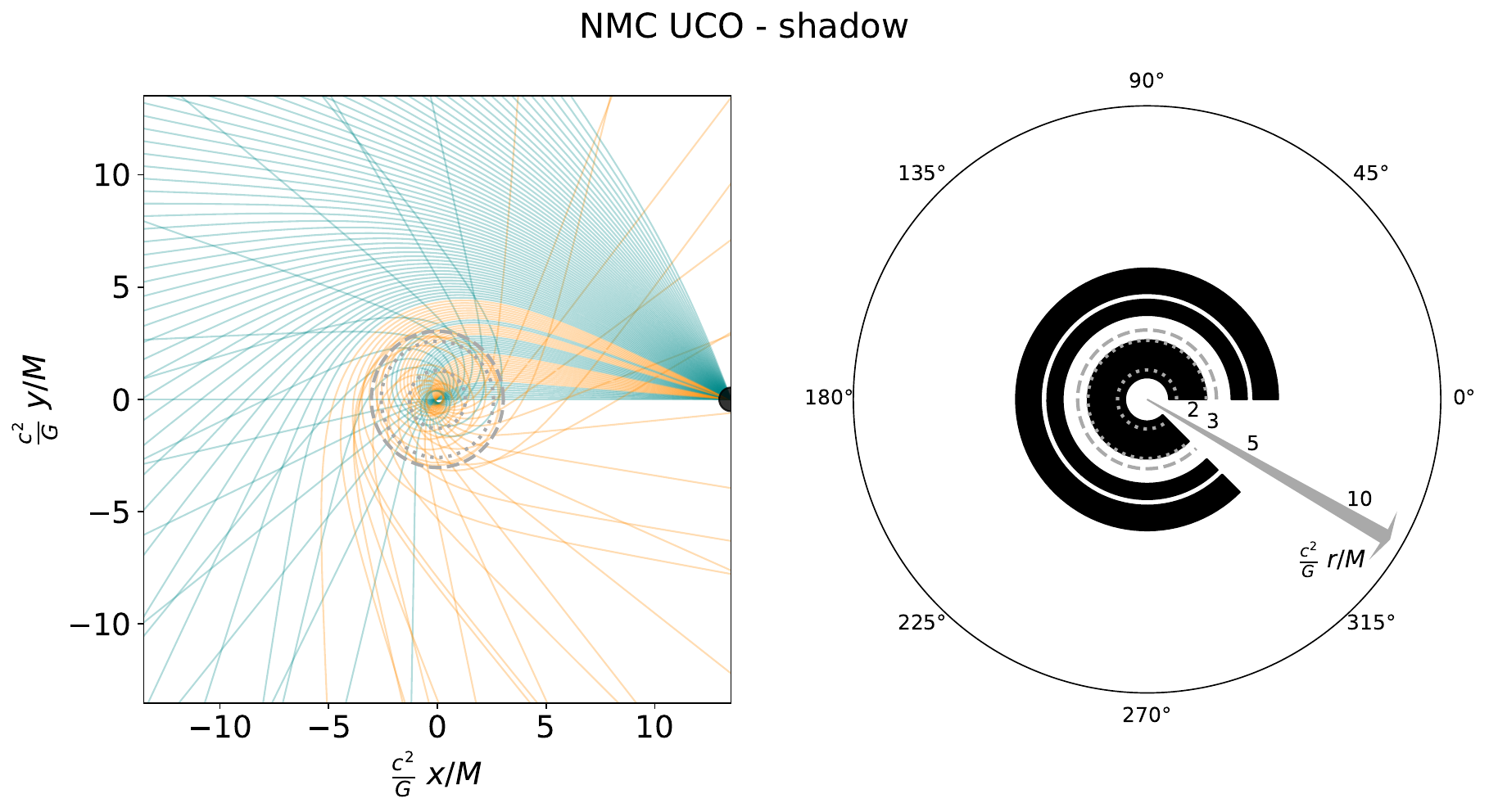}
    \caption{Pseudo-shadow of the UCO solution with central density $\hat{\rho}_0=0.95$. Left panel illustrates a ray-tracing simulation, with the observer at the black spot on the right and an illuminated screen assumed at $x=-\infty$. Solid lines are light rays traced back in time from the observer with different impact parameters $j/E$: in the remote past, cyan rays hit the illuminated screen at $x=-\infty$, and hence can be seen by the observer; orange rays are instead deflected back to $x=+\infty$ and are precluded to vision. Dashed grey lines displays the position of the boundary, dotted grey lines that of the inner and outer light rings. Right panel is a polar plot illustrating the face-on view of the UCO from the observer position: as a consequence of gravitational lensing, a multiple photon ring structure is originated  around the typical radii $(c^2/G)\, r\approx 1.8-4-5\, M$.}
    \label{fig|shadow}
\end{figure}

A century after the pioneering observation of gravitational light deflection during a solar eclipse that provided the first smoking-gun test of General Relativity, in 2019 the Event Horizon Telescope (EHT) Collaboration published the first "image" of a BH \cite{EHTI}. In fact, the light rays passing close to a BH can be deflected very strongly and even travel on circular orbits in correspondence of the light rings. This strong deflection, together with the fact that no light comes out of a BH, allows the latter to be imaged as a dark disk in the sky, known as "BH shadow". For a Schwarzschild black hole the size of the shadow is $r_{\rm sh} = 3\sqrt{3}\, M\approx 5.2\, M$ (e.g., see \cite{Perlick_2022} for a derivation and an extended discussion on black holes' shadows). 

The NMC-UCOs do not feature an horizon and, being composed of DM, are completely transparent to light rays; nevertheless, the notion of a shadow in a wide sense can be introduced also for them. The left panel of Figure \ref{fig|shadow} shows a ray-tracing simulation, with the observer positioned on the black spot at the extreme right of the plot and an illuminated screen assumed at $x=-\infty$. Solid lines represent light rays from the observer traced back in time for different impact parameters $j/E$. Light rays stemming from the observer are divided in two classes: those that in the remote past hit the illuminated screen at $x=-\infty$ can be seen by the observer and are highlighted in cyan; orange rays are instead deflected back to $x=+\infty$ and not reaching the illuminated screen are precluded to vision. Right panel is a polar plot illustrating the face-on view of the UCO from the observer position. This originates a pseudo-shadow at $\approx 5.2\,M$, close to the size of that for a Schwarzschild black hole. Inside it, a multiple photon ring structure around the typical radii $(c^2/G)\, r\approx 1.8-4-5\, M$ emerges, as generically expected for horizonless compact objects \cite{carballo2024}. 

\section{Discussion and conclusions}\label{sec|discussion}

In this work we have presented a general relativistic action featuring a non-minimal coupling between the stress energy tensor of a collisionless dark matter fluid and gravity. The latter could be described in terms of a scattering cross section between dark matter particles and gravitons which, if sufficiently small, opens the intriguing possibility that the non-minimal coupling with gravity could be present also for standard model particles, but that for the latter its strength would be outclassed by the other interactions.

We have shown that the non-minimal coupling originates an anisotropic effective pressure able to counteract gravity and to support the otherwise collisionless dark matter fluid into static, spherically symmetric configurations.
Investigating the properties of such configurations, we have found that they are horizonless, non singular ultra-compact objects (NMC-UCOs), with a very limited range of allowed central densities. We have also analyzed the orbits of massive and massless test particles in the gravitational field of NMC-UCOs, providing some specific examples of both bound and unbound orbits and a general discussion in terms of phase portraits. Finally, we have implemented a basic ray-tracing simulation to study the gravitational lensing effects around NMC-UCOs. We have effectively described these in terms of a pseudo-shadow that makes connection with what is usually done in the context of black holes. If observed, the peculiar light pattern around NMC-UCOs could constitute a clear signature for the detection of such exotic objects.

A caveat concerns the densities required to form NMC-UCOs, that ultimately depend on the magnitude of the lengthscale parameter $L$ appearing in the relativistic theory; in particular for $L \sim 10^{10} \,\rm cm$ one finds densities $\rho \sim 10^7 \,\rm g/cm^3$ and masses of astrophysical interest $M \sim 10^4 \,\rm M_{\odot}$. However, such high densities, necessary to make the non-minimal coupling to kick in, are considerably higher that the average ones occurring within virialized dark matter halos. Nevertheless, the possibility remains that the necessary conditions could have been met during the early fast and violent collapse of a dark matter halo or in the early Universe. In the latter instance, the environmental conditions could be prone for a copious production of horizonless NMC-UCOs that, at variance with primordial black holes, will not suffer evaporation-related issues and thus may potentially survive till much later cosmic times. In a forthcoming paper we plan to study the collapse of NMC-UCOs, and to investigate whether they could have some astrophysical relevance as primordial gravitational wells for the early formation of supermassive black hole seeds.

\section*{Acknowledgements}
We warmly thank Bruce Bassett and Giovanni Gandolfi for illuminating discussions. This work was partially funded from the projects: "Data Science methods for MultiMessenger Astrophysics \& Multi-Survey Cosmology" funded by the Italian Ministry of University and Research, Programmazione triennale 2021/2023 (DM n.2503 dd. 9 December 2019), Programma Congiunto Scuole; EU H2020-MSCA-ITN-2019 n. 860744 \textit{BiD4BESt: Big Data applications for black hole Evolution STudies}; Italian Research Center on High Performance Computing Big Data and Quantum Computing (ICSC), project funded by European Union - NextGenerationEU - and National Recovery and Resilience Plan (NRRP) - Mission 4 Component 2 within the activities of Spoke 3 (Astrophysics and Cosmos Observations);  European Union - NextGenerationEU under the PRIN MUR 2022 project n. 20224JR28W "Charting unexplored avenues in Dark Matter"; INAF Large Grant 2022 funding scheme with the project "MeerKAT and LOFAR Team up: a Unique Radio Window on Galaxy/AGN co-Evolution; INAF GO-GTO Normal 2023 funding scheme with the project "Serendipitous H-ATLAS-fields Observations of Radio Extragalactic Sources (SHORES)"

\bibliographystyle{unsrt}
\bibliography{bibliography}

\end{document}